\documentclass[aps,prc,preprint,amsmath,amssymb,showpacs,preprintnumbers,superscriptaddress]{revtex4-1}
\usepackage{CJK}
\usepackage{graphicx}
\usepackage{dcolumn}
\usepackage{bm}
\usepackage{color}
\usepackage{hyperref}
\allowdisplaybreaks[4]

\begin{document}
\begin{CJK*}{UTF8}{}

  \title{Dipole response in deformed halo nuclei $^{42}\mathrm{Mg}$ and $^{44}\mathrm{Mg}$}
  \author{X. F. Jiang \CJKfamily{gbsn} (姜晓飞)}
  \affiliation{State Key Laboratory of Nuclear Physics and Technology, School of Physics, Peking University, Beijing 100871, China}

  \author{Z. Z. Li \CJKfamily{gbsn} (李征征)}
  \affiliation{State Key Laboratory of Nuclear Physics and Technology, School of Physics, Peking University, Beijing 100871, China}

  \author{X. W. Sun \CJKfamily{gbsn} (孙旭伟)}
  \affiliation{School of Physics, Engineering and Technology, University of York, Heslington, York YO10 5DD, United Kingdom}

  \author{J. Meng \CJKfamily{gbsn} (孟杰)}
  \email{mengj@pku.edu.cn}
  \affiliation{State Key Laboratory of Nuclear Physics and Technology, School of Physics, Peking University, Beijing 100871, China}
  \affiliation{Center for Theoretical Physics, China Institute of Atomic Energy, Beijing 102413, China}

  \begin{abstract}
The quasiparticle finite amplitude method based on the deformed relativistic Hartree-Bogoliubov theory in continuum has been developed for the noncharge-exchange multipole response.
Taking neutron-rich magnesium isotopes as examples, the isovector electric dipole response, especially in the low-lying region, is studied.
It is found that the low-energy dipole strength increases with neutron number and becomes notably enhanced in the predicted deformed halo nuclei $^{42}\mathrm{Mg}$ and $^{44}\mathrm{Mg}$.
In these isotopes, the $K^\pi=1^-$ states below 3 MeV are dominated by transitions from the ``halo" part of the single-neutron orbitals.
Their transition densities reveal a low-frequency, out-of-phase oscillation between the neutron halo and the core.
These results provide a microscopic picture for the soft dipole resonance in $^{42}\mathrm{Mg}$ and $^{44}\mathrm{Mg}$.
  \end{abstract}


  \maketitle

\end{CJK*}

\section{Introduction}\label{Section1}

New phenomena discovered in the exotic nuclei far from the $\beta$-stability line, such as the halo~\cite{Tanihata1985PRL}, new magic numbers~\cite{Ozawa2000PRL}, island of inversion~\cite{Warburton1990PRC}, and pygmy resonances~\cite{Adrich2005PRL}, have promoted the construction and upgrade of the radioactive ion beam facilities~\cite{ZhouXH2022AAPPSBulletin,YuanYJ2020JPGCS,Castelvecchi2022Nature,Sakurai2018FoP,Durante2019PS,Hong2023JPCS,Ball2011JPGNPP}.
These phenomena also provide challenges to nuclear theory developed from the $\beta$-stable nuclei.

During the past decades, the relativistic density functional theory (DFT) has been proven to be powerful in successfully describing various nuclear properties and predicting new phenomena~\cite{Mengbook}.
In order to describe the exotic nuclei, by taking into account the pairing correlations and continuum effects in a microscopic and self-consistent way, the relativistic continuum Hartree-Bogoliubov (RCHB) theory was developed~\cite{Meng1996PRL,Meng1998NPA}.
The RCHB theory successfully describes the ground states of the spherical nuclei both near and far from the $\beta$-stability line~\cite{Meng2006PPNP}, interprets the halo in $^{11}\mathrm{Li}$~\cite{Meng1996PRL}, and predicts the giant halo in Zr~\cite{Meng1998PRL} and Ca~\cite{Meng2002PRC} isotopes.
To describe the deformed exotic nuclei, the deformed relativistic Hartree-Bogoliubov theory in continuum (DRHBc) was developed by solving the relativistic Hartree-Bogoliubov (RHB) equations in a Dirac Woods-Saxon basis~\cite{ZhouSG2010PRC,LiLL2012PRC}.
Inheriting the advantages of the RCHB theory and including the deformation degrees of freedom, the DRHBc theory has been applied to study deformed exotic nuclei, including $^{17,19}\mathrm{B}$~\cite{YangZH2021PRL,Sun2021PRCa}, $^{22}\mathrm{C}$~\cite{SunXX2018PLB}, $^{31}\mathrm{Ne}$~\cite{ZhongSY2022SCPMA,PanC2024PLB}, $^{37}\mathrm{Mg}$~\cite{ZhangKY2023PLB}, and to predict the halos in $^{39}\mathrm{Na}$~\cite{ZhangKY2023PRC} and $^{42,44}\mathrm{Mg}$~\cite{ZhouSG2010PRC,LiLL2012PRC,ZhangKY2019PRC,SunXX2021SB}.

The random phase approximation (RPA) method based on the DFT has been widely used to describe the nuclear collective excitations~\cite{ManyBody}.
However, for deformed exotic nuclei, the configuration space becomes very large, and the calculation and diagonalization of the RPA matrix can be extremely time-consuming.
As an alternative, the finite amplitude method (FAM)~\cite{Nakatsukasa2007PRC}, which iteratively solves the linear response of the fields induced by the one-body transition operator, has been proven to be efficient for deformed nuclei.
The FAM has been implemented based on both nonrelativistic~\cite{Inakura2009PRC,Avogadro2011PRC,Hinohara2013PRC,Mustonen2014PRC,PeiJC2014PRC} and relativistic DFTs~\cite{LiangHZ2013PRC,Niksic2013PRC,SunXW2017PRC,Bjelcic2020CPC,Bjelcic2023CPC,ZhaoJ2024PRC,ChenC2025PRC}, and applied to study the soft monopole mode~\cite{PeiJC2014PRC,SunXW2021PRC}, pygmy dipole resonance~\cite{Inakura2011PRC}, $\beta$-decay half-lives~\cite{Mustonen2014PRC,Shafer2016PRC}, the collective inertia in spontaneous fission~\cite{Washiyama2021PRC}, and the $\gamma$-ray strength functions~\cite{GonzalezMiretZaragoza2025PRC}, etc.

Exotic excitation modes of nuclei far from the $\beta$-stability line~\cite{Paar2007RPP} reveal how nuclear matter behaves under extreme conditions.
For the electric dipole (E1) response in neutron-rich nuclei, the pygmy dipole resonance (PDR) is the low-energy mode corresponding to the resonant oscillation of the weakly bound neutron skin against the isospin-saturated proton-neutron core~\cite{Paar2007RPP}.
The PDR provides constrains on the symmetry energy parameters~\cite{Carbone2010PRC,Vretenar2012PRC}, significantly influences the neutron capture rates in the \emph{r}-process nucleosynthesis~\cite{Brzosko1969CJP,Goriely1998PLB,Berceanu2021PRC}, and thus has attracted wide attention~\cite{Paar2007RPP,Savran2013PPNP,Aumann2019EPJA,Bracco2019PPNP,Lanza2023PPNP}.
Apart from the PDR, a new type of low-energy E1 response may appear in halo nuclei~\cite{Aumann2013PS,Tanihata2019EPJA,Aumann2019EPJA,Nakamura2023Book,LuX2025PRC}.
For example, in the halo nuclei $^{11}\mathrm{Li}$, which is interpreted as a dineutron coupled with a nuclear core $^{9}\mathrm{Li}$, is predicted to have a very soft dipole mode~\cite{Hansen1987EpL}.
The predicted large cross section for Coulomb dissociation corresponding to this soft dipole mode was later verified experimentally~\cite{Kobayashi1989PLB}.
The soft dipole mode, namely the enhanced low-lying E1 strength in halo nuclei, triggered the conjecture of the soft dipole resonance, a new type of collective oscillational motion characterized by low-frequency oscillations between the halo and the core~\cite{Ikeda1992NPA}.
This conjecture has attracted extensive theoretical studies, including the cluster-orbital shell model~\cite{Suzuki1990NPA,Myo2022PRC,Myo2022PTEP,Myo2023PRC}, the nuclear shell model~\cite{Suzuki2000NPA}, the continuum-discretized coupled-channel method~\cite{Matsumoto2019PTEP}, and the Gamow coupled-channel method~\cite{YangYH2025EPJA}.
Experimentally, a strong transition amplitude indicating the soft-resonance of $^{11}\mathrm{Li}$ has been observed in $(d,d^\prime)$~\cite{Kanungo2015PRL} and $(p,p^\prime)$~\cite{Tanaka2017PLB} inelastic scattering.

The combination of DRHBc with the FAM framework offers a powerful tool for investigating collective excitations in exotic nuclei.
Based on the DRHBc theory, the FAM has been implemented to study the isoscalar giant monopole resonance in exotic nuclei~\cite{SunXW2022PRCa}.

In this work, we have developed the quasiparticle finite amplitude method (QFAM) based on the DRHBc theory for multipole excitations, by explicitly linearizing the Dirac Hamiltonian and pairing potential.
As an example, the isovector electric dipole responses in neutron-rich even-even magnesium isotopes $^{28-44}\mathrm{Mg}$ are studied.
The strength functions of $K^\pi=0^-$ and $K^\pi=1^-$ excitations are discussed and compared.
Our focus is placed specifically on the predicted deformed halo nuclei $^{42}\mathrm{Mg}$ and $^{44}\mathrm{Mg}$ to explore the soft dipole resonance.

\section{Theoretical Framework}\label{Section2}

\subsection{Deformed relativistic Hartree-Bogoliubov theory in continuum}\label{Subsection2.1}

The detailed formalism of the DRHBc theory can be seen in Refs.~\cite{LiLL2012PRC,ZhangKY2020PRC}.
In the DRHBc theory, the mean field and pairing correlations are treated self-consistently by the RHB equation,
\begin{equation}
  \left(\begin{array}{c c}{{\hat{h}_{D}-\lambda_\tau}}&{{\hat{\Delta}}}\\ {{-\hat{\Delta}^{\ast}}}&{{-\hat{h}_{D}^{\ast}+\lambda_\tau}}\end{array}\right)\left(\begin{array}{c}{{U_{k}}}\\ {{V_{k}}}\end{array}\right)=E_{k}\left(\begin{array}{c}{{U_{k}}}\\ {{V_{k}}}\end{array}\right),\label{Eq1}
\end{equation}
where $\hat{h}_{D}$ is the Dirac Hamiltonian, $\lambda_\tau$ is the Fermi energy ($\tau=n/p$ for neutrons and protons), $\hat{\Delta}$ is the pairing potential, $E_k$ is the quasiparticle energy, and $U_k$ and $V_k$ are the quasiparticle wavefunctions.
In coordinate space, the Dirac Hamiltonian $\hat{h}_{D}$ reads
\begin{equation}
  \hat{h}_{D}(\bm r)=\bm\alpha\cdot(\bm p-\bm V)+V^0+\beta(M+S),\label{Eq2}
\end{equation}
where $S$ is the scalar potential, and $V^0$ and $\bm V$ are the time and space components of the vector potential $V^\mu$, respectively.
In the ground states of even-even nuclei, the vector potential $\bm V$ vanishes due to time-reversal symmetry.
In the framework of point-coupling density functional~\cite{Mengbook},
\begin{eqnarray}
  S(\bm r) & = & \alpha_{S}\rho_{S}+\beta_{S}\rho_{S}^{2}+\gamma_{S}\rho_{S}^{3}+\delta_{S}\Delta\rho_{S},\label{Eq3} \\
  V^\mu(\bm r) & = & \alpha_{V}j^\mu+\gamma_{V}(j_\nu j^\nu)j^\mu+\delta_{V}\Delta j^\mu+\tau_{3}\alpha_{TV}j_3^\mu+\tau_{3}\delta_{TV}\Delta j_3^\mu+e\frac{1-\tau_3}{2}A^\mu,\label{Eq4}
\end{eqnarray}
where $\rho_S$, $j^\mu$, and $j_3^\mu$ are scalar density, vector current and isovector current, respectively.
The pairing potential $\hat\Delta$ reads
\begin{equation}
  \hat\Delta(\bm r_1,\bm r_2)=V^{pp}(\bm r_1,\bm r_2)\kappa(\bm r_1,\bm r_2),\label{Eq5}
\end{equation}
where $\kappa$ is the pairing tensor and $V^{pp}$ is a density-dependent zero-range pairing force,
\begin{equation}
  V^{pp}(\bm r_1,\bm r_2)=V_0\frac12(1-P^\sigma)\delta(\bm r_1-\bm r_2)\left(1-\frac{\rho^\tau(\bm r_1)}{\rho_{\mathrm{sat}}}\right).\label{Eq6}
\end{equation}

In DRHBc theory, the RHB equation~\eqref{Eq1} is solved in a spherical Dirac Woods-Saxon (DWS) basis~\cite{ZhouSG2003PRC,ZhangKY2022PRC}.
As the wave functions of the DWS basis have a proper asymptotic behavior at large radius $r$, the DRHBc theory is capable of including the contributions from the continuum and describing the exotic nuclei with large spatial extensions.
In order to describe axially deformed nuclei with spatial reflection symmetry, the potentials and densities are expanded in terms of the Legendre polynomials
\begin{equation}
  f(\bm r)=\sum_{\lambda}f_\lambda(r)P_\lambda(\cos\theta),\quad\lambda=0,2,4,\dots,\lambda_\mathrm{max}\label{Eq7}
\end{equation}
with
\begin{equation}
  f_\lambda(r)=\frac{2\lambda+1}{4\pi}\int d\Omega f(\bm r)P_\lambda(\cos\theta).\label{Eq8}
\end{equation}

\subsection{Quasi-particle finite amplitude method}\label{Subsection2.2}

The detailed formalism of the QFAM can be seen in Refs.~\cite{Avogadro2011PRC,Niksic2013PRC,Bjelcic2020CPC}.
The response of a nucleus under a time-dependent external field is determined by the time-dependent relativistic Hartree-Bogoliubov (TDRHB) equation
\begin{equation}
  \mathrm{i}\hbar\frac{\partial}{\partial t}\mathcal{R}(t)=[\mathcal{H}(\mathcal{R}(t))+\mathcal{F}(t),\mathcal{R}(t)].\label{Eq9}
\end{equation}
where $\mathcal{R}(t)$ is the generalized density, $\mathcal{H}(t)$ is the quasiparticle Hamiltonian, and $\mathcal{F}(t)$ is the external field.
For a weak harmonic external field with the frequency $\omega$
\begin{equation}
  \mathcal{F}(t)=\mathcal{F}e^{-\mathrm{i}\omega t}+\mathcal{F}^\dagger e^{\mathrm{i}\omega t},\label{Eq10}
\end{equation}
in the small amplitude limit, $\mathcal{R}(t)$ and $\mathcal{H}(t)$ oscillate around the stationary density $\mathcal{R}_0$ and Hamiltonian $\mathcal{H}_0$ with the same frequency, respectively
\begin{eqnarray}
  \mathcal{R}(t)&=&\mathcal{R}_0+\delta\mathcal{R}(\omega)e^{-\mathrm{i}\omega t}+\delta\mathcal{R}^\dagger(\omega)e^{\mathrm{i}\omega t},\label{Eq11}\\
  \mathcal{H}(t)&=&\mathcal{H}_0+\delta\mathcal{H}(\omega)e^{-\mathrm{i}\omega t}+\delta\mathcal{H}^\dagger(\omega)e^{\mathrm{i}\omega t}.\label{Eq12}
\end{eqnarray}
By inserting the Eqs.~\eqref{Eq10} to~\eqref{Eq12} into Eq.~\eqref{Eq9} and expanding up to the linear order in the external field, the linear response equation is obtained, which, in the quasiparticle basis, takes the form
\begin{eqnarray}
    (E_\mu+E_\nu-\omega)X_{\mu\nu}(\omega)+\delta H^{20}_{\mu\nu}(\omega)&=-F^{20}_{\mu\nu},\label{Eq13}\\
    (E_\mu+E_\nu+\omega)Y_{\mu\nu}(\omega)+\delta H^{02}_{\mu\nu}(\omega)&=-F^{02}_{\mu\nu}.\label{Eq14}
\end{eqnarray}
Here, $E_\mu$ and $E_\nu$ are quasiparticle energies, and $F^{20}_{\mu\nu}$ and $F^{02}_{\mu\nu}$ are matrix elements of the two-quasiparticle creation and annihilation parts of the external field, respectively.
The induced Hamiltonian $\delta H^{20}(\omega)$ and $\delta H^{02}(\omega)$ are functionals of the QFAM amplitudes $X(\omega)$ and $Y(\omega)$, which are defined as
\begin{equation}
  \delta \mathcal{R}(\omega)
  =\left(\begin{array}{cc}0&\delta R^{20}(\omega)\\-\delta R^{02}(\omega)&0\end{array}\right)
  \equiv\left(\begin{array}{cc}0&X(\omega)\\-Y(\omega)&0\end{array}\right).\label{Eq15}
\end{equation}

In the QFAM, the linear response equation~\eqref{Eq13} and~\eqref{Eq14} are solved iteratively.
This iterative process involves calculating the induced Dirac Hamiltonian $\delta h_D$ and pairing potential $\delta\Delta$ from the induced single-particle density $\delta\rho$ and pairing tensor $\delta\kappa$.
In Refs.~\cite{Avogadro2011PRC,Niksic2013PRC,SunXW2022PRCa}, the Dirac Hamiltonian and pairing potential are respectively calculated for both the stationary state $(\rho,\kappa)$ and perturbed state $(\rho+\delta\rho,\kappa+\delta\kappa)$, and the differences give the perturbed quantities.
The advantage of this method is that the QFAM can be implemented by a little extension of the original stationary code.
The disadvantage is that the induced densities and potentials are constrained by the symmetry assumptions of the DRHBc theory, which limits the application for nuclear excitation modes with $K^\pi=0^+$ only.

In the present work, the induced Dirac Hamiltonian is calculated from the explicit linearization of the Dirac Hamiltonian in Eq.~\eqref{Eq2}
\begin{equation}
  \delta h_D(\bm r)=-\bm\alpha\cdot\delta\bm V+\delta V^0+\beta\delta S,\label{Eq16}
\end{equation}
where the induced potentials read
\begin{eqnarray}
  \delta S(\bm r)&=&\left[\alpha_S+2\beta_S\rho_S+3\gamma_V(\rho_S)^2\right]\delta\rho_S+\delta_S\triangle\delta\rho_S,\label{Eq17} \\
  \delta V^0(\bm r)&=&\left[\alpha_V+3\gamma_V(j_0j^0)\right]\delta j^0+\delta_V\triangle\delta j^0+\tau_3\alpha_{TV}\delta j_3^0+\tau_3\delta_{TV}\triangle\delta j_3^0+e\frac{1-\tau_3}{2}\delta A^0,\label{Eq18} \\
  \delta\bm V(\bm r)&=&\left[\alpha_V+\gamma_V(j_0j^0)\right]\delta \bm j+\delta_V\triangle\delta\bm j+\tau_3\alpha_{TV}\delta\bm j_3+\tau_3\delta_{TV}\triangle\delta\bm j_3+e\frac{1-\tau_3}{2}\delta\bm A.\label{Eq19}
\end{eqnarray}
The induced pairing potential reads
\begin{equation}
  \delta\Delta(\bm r_1,\bm r_2)=V^{pp}(\bm r_1,\bm r_2)\delta\kappa(\bm r_1,\bm r_2).\label{Eq20}
\end{equation}
where the variation of the pairing interaction $\delta V^{pp}$ is neglected.
In order to describe induced densities and potentials without any symmetry assumption, the time and space components are expanded using scalar and vector spherical harmonics~\cite{Varshalovich1988book}, respectively.

In practice, a small imaginary part $\gamma$ is added to the frequency, $\omega_\gamma=\omega+\mathrm{i}\gamma$, which corresponds to a Lorentzian smearing width $2\gamma$.
By solving the linear response equation at the complex frequency $\omega_\gamma$, the QFAM amplitudes $X(\omega_\gamma)$ and $Y(\omega_\gamma)$ are obtained, and are then used to calculate the strength function,
\begin{equation}
  S(\omega)=-\frac{1}{\pi}\mathrm{Im}\left[\frac12\sum_{\mu\nu}F^{20*}_{\mu\nu}X_{\mu\nu}(\omega_\gamma)+F^{02*}_{\mu\nu}Y_{\mu\nu}(\omega_\gamma)\right].\label{Eq21}
\end{equation}
The present work is concentrated on the isovector dipole response induced by the operator
\begin{equation}
  \hat{F}_{1K}=e\frac{N}{A}\sum_{i=1}^{Z}r_iY_{1K}(\hat{r}_i)-e\frac{Z}{A}\sum_{i=1}^{N}r_iY_{1K}(\hat{r}_i).\label{Eq22}
\end{equation}

\section{Numerical details}\label{Section3}

In the DRHBc calculations, the relativistic density functional PC-PK1~\cite{Zhao2010PRC} is employed in the particle-hole channel.
In the particle-particle channel, the density-dependent zero-range pairing force in Eq.~\eqref{Eq6} is adopted, with the pairing strength $V_0=-325\ \mathrm{MeV\cdot fm^3}$, the saturation density $\rho_\mathrm{sat}=0.152\ \mathrm{fm^{-3}}$, and a pairing window of 100 MeV, following the DRHBc mass table~\cite{ZhangKY2020PRC,Pan2022PRC,Zhang2022ADNDT,Guo2024ADNDT}.
The energy cutoff for the DWS basis in the Fermi sea is $E^+_\mathrm{cut}=150\ \mathrm{MeV}$.
The number of DWS basis states in the Dirac sea is taken to be the same as that in the Fermi sea.
The angular momentum cutoff for the DWS basis is $J_\mathrm{cut}=23/2\hbar$.
The Legendre expansion truncation in Eq.~\eqref{Eq7} is $\lambda_\mathrm{max}=6$.

In the QFAM calculations, the same numerical details are used.
A full two-quasiparticle configuration space is constructed without any additional truncation.
The truncation of both scalar and vector spherical harmonics expansion~\cite{Varshalovich1988book} of induced densities and potentials is $L_\mathrm{max}=6$.

\section{Results and discussion}\label{Section4}

The ground states of neutron-rich even-even magnesium isotopes $^{28-44}\mathrm{Mg}$ are calculated by the DRHBc theory with PC-PK1 functional.
As has been found in Refs.~\cite{Zhang2022ADNDT,Guo2024ADNDT}, the Mg isotopes under investigation are prolately deformed, except for the spherical $^{32}\mathrm{Mg}$.
Among them, $^{42,44}\mathrm{Mg}$ are predicted to be deformed halo nuclei by different density functionals~\cite{ZhouSG2010PRC,LiLL2012PRC}.
In the following, a systematic QFAM calculation based on DRHBc is carried out for the isovector electric dipole strength functions in these isotopes.

\begin{figure}[htbp]
  \centering
  \includegraphics[width=0.8\textwidth]{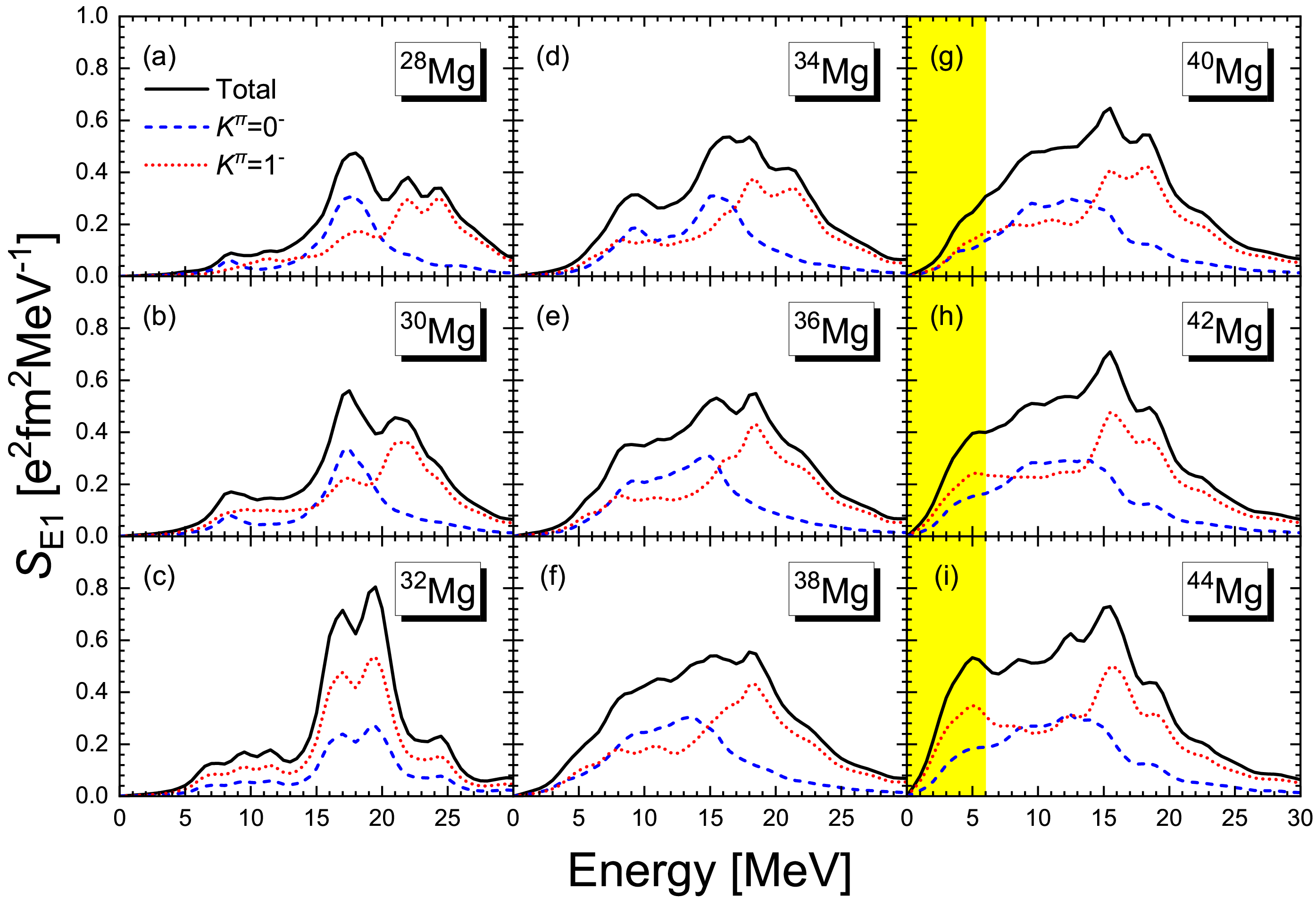}
  \caption{Isovector electric dipole strength functions in neutron-rich even-even magnesium isotopes, calculated by QFAM with a Lorentzian smearing width of 2 MeV. The dashed, dotted, and solid lines correspond to the $K^\pi=0^-$, $K^\pi=1^-$, and total strength functions, respectively. For $K^\pi=1^-$ response, the transition strength for the $K^\pi=\pm1^-$ excitations are summed up. The low-energy regions (0-6 MeV) in $^{40-44}\mathrm{Mg}$ are highlighted.}
  \label{fig1}
\end{figure}

Fig.~\ref{fig1} shows the isovector electric dipole strength functions in neutron-rich even-even magnesium isotopes $^{28-44}\mathrm{Mg}$, with a Lorentzian smearing width of 2 MeV.
The $K^\pi=0^-$ and $K^\pi=1^-$ giant dipole resonances correspond to the collective oscillations of protons against neutrons along and perpendicular to the symmetry axis, respectively.
The peak energies of $K^\pi=0^-$ and $K^\pi=1^-$ strength functions are different except for the ones in the spherical $^{32}\mathrm{Mg}$.
As shown in Fig.~\ref{fig1}, the prolate deformation causes the $K^\pi=0^-$ strength function to be red-shifted because the restoring force scales inversely with the elongation of the symmetry axis.
From $^{28}\mathrm{Mg}$ to $^{40}\mathrm{Mg}$, for energies below 6 MeV, no significant discrepancy between $K^\pi=0^-$ and $K^\pi=1^-$ strength functions can be observed.
In $^{42}\mathrm{Mg}$ and $^{44}\mathrm{Mg}$, the strength for $K^\pi=1^-$ is much more enhanced than that of $K^\pi=0^-$.
The dipole strength below 6 MeV in $^{40-44}\mathrm{Mg}$ will be discussed in the following.
It is also noticed, our calculations based on the PC-PK1 functional predict more fragmented distributions compared with the relatively sharp strength distributions in $^{36,38,40}\mathrm{Mg}$ based on SkM* functional~\cite{Yoshida2009PRC} and in $^{40}\mathrm{Mg}$ based on the extended SLy4 functional~\cite{WangK2017PRC}, which may be related to the difference in the slope parameter of symmetry energy $L$~\cite{Carbone2010PRC}.

\begin{figure}[!htbp]
  \centering
  \includegraphics[width=0.8\textwidth]{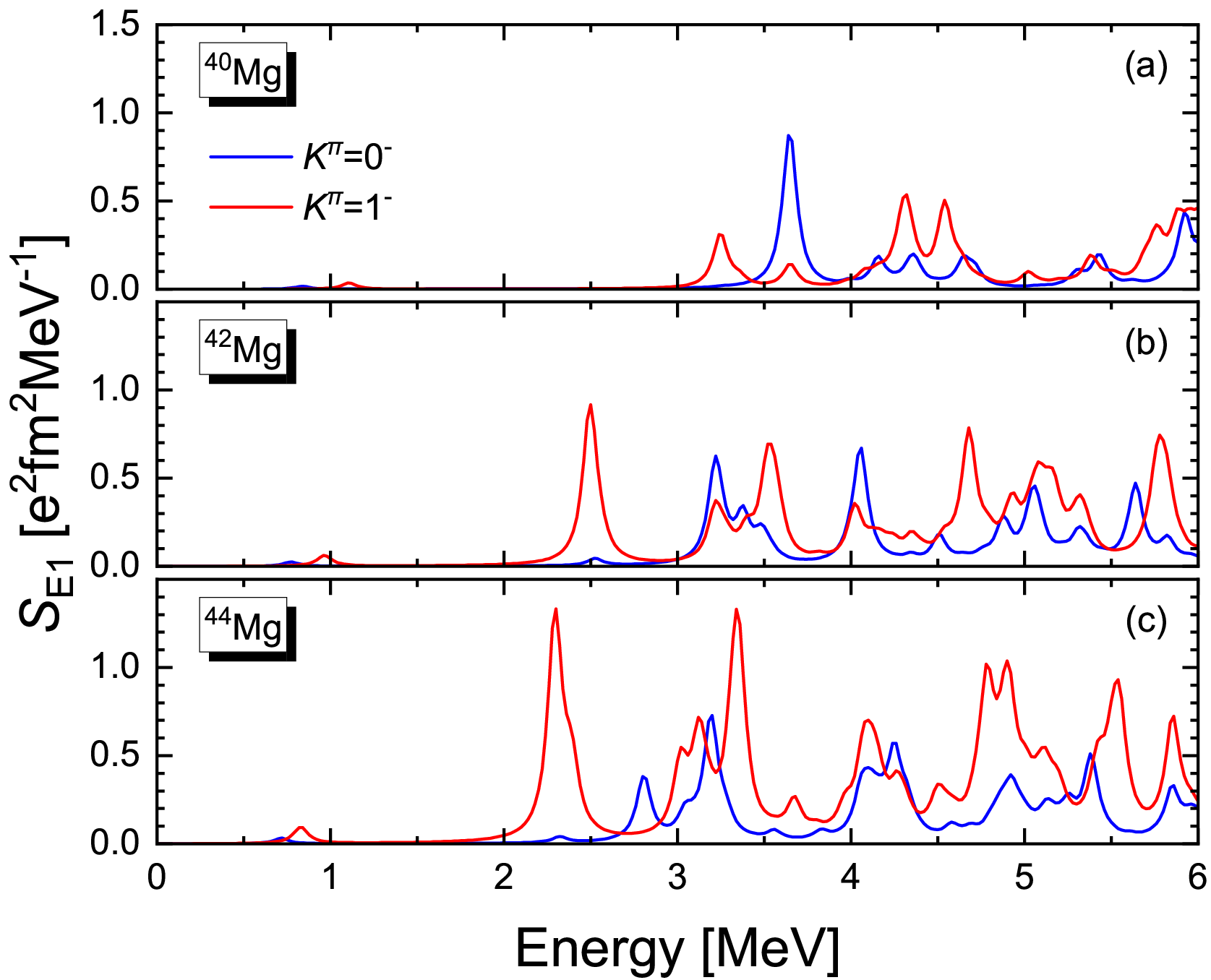}
  \caption{Isovector electric dipole strength function below 6 MeV in $^{40}\mathrm{Mg}$ (a), $^{42}\mathrm{Mg}$ (b), and $^{44}\mathrm{Mg}$ (c), calculated by QFAM with a Lorentzian smearing width of 0.1 MeV. The $K^\pi=0^-$ and $K^\pi=1^-$ strength functions are denoted by blue and red lines, respectively.}
  \label{fig2}
\end{figure}

The isovector electric dipole strength functions below 6 MeV in $^{40,42,44}\mathrm{Mg}$, calculated by QFAM with a Lorentzian smearing width of 0.1 MeV, are shown in Fig.~\ref{fig2}.
Apart from the fragmented strength functions between 3 and 6 MeV, the strength functions show additional $K^\pi=1^-$ peaks around 2.5 MeV in $^{42}\mathrm{Mg}$, and around 2.3 MeV in $^{44}\mathrm{Mg}$.
Further analysis with a smearing width of 0.002 MeV reveals that the peak in $^{42}\mathrm{Mg}$ is contributed by two $K^\pi=1^-$ states at energies 2.50 MeV and 2.56 MeV. 
Similarly, in $^{44}\mathrm{Mg}$ the peak is contributed by two $K^\pi=1^-$ states at 2.30 MeV and 2.39 MeV.

\begin{figure*}[!htbp]
  \centering
  \includegraphics[width=0.8\textwidth]{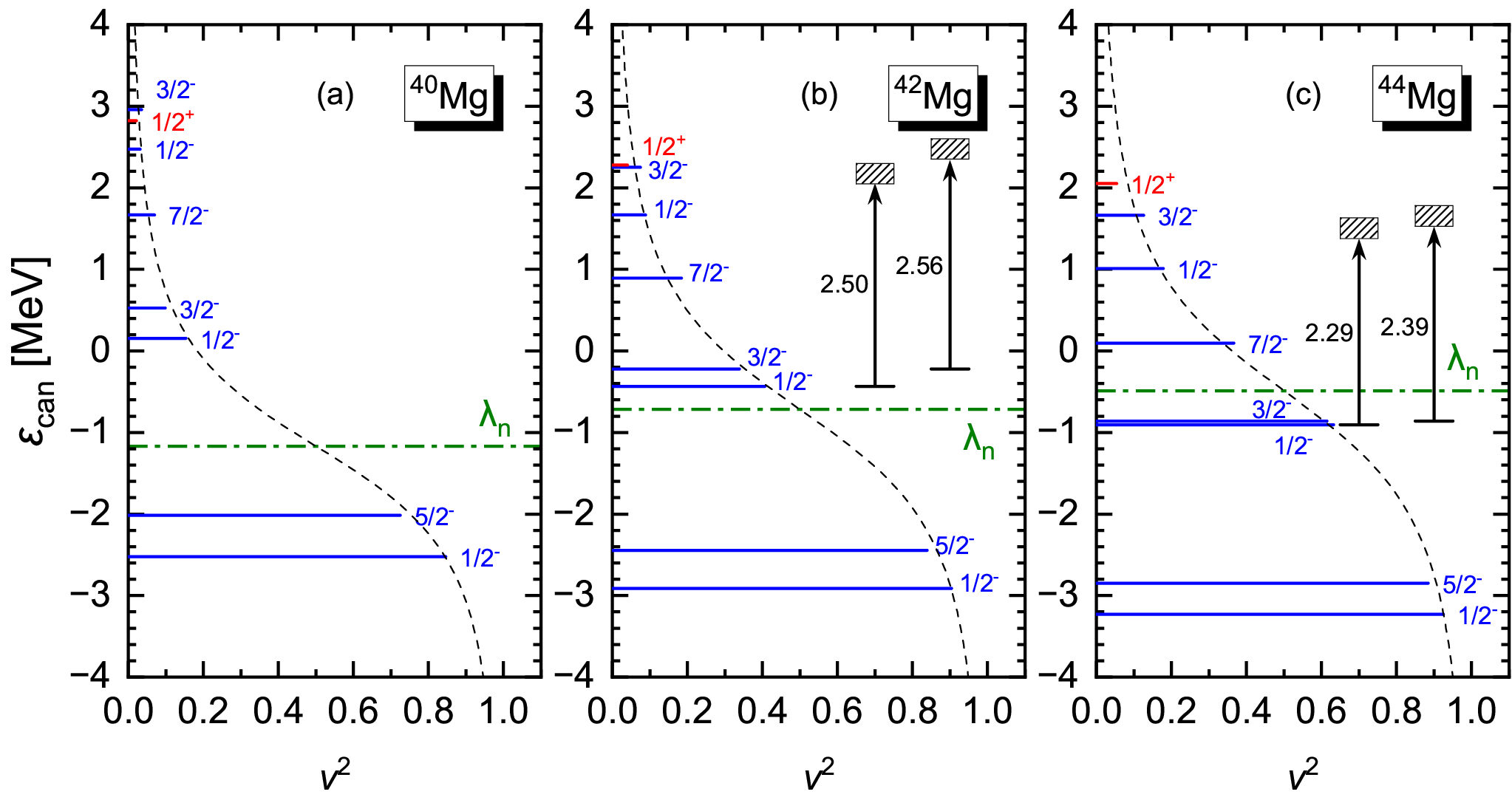}
  \caption{The single-neutron levels around the Fermi surface in the canonical basis for $^{40}\mathrm{Mg}$ (a), $^{42}\mathrm{Mg}$ (b), and $^{44}\mathrm{Mg}$ (c) versus the occupation probability. The levels are labeled by the good quantum numbers of parity $\pi$ (the superscript) and the third component of angular momentum $m$. The dash-dotted lines represent the neutron Fermi surfaces. The occupation probabilities derived from the BCS formula with the corresponding average pairing gap are shown by the dashed curves. The dominant excitations for $K^\pi=1^-$ states below 3 MeV in $^{42}\mathrm{Mg}$ and $^{44}\mathrm{Mg}$ are labeled schematically.}
  \label{fig3}
\end{figure*}

Fig.~\ref{fig3} shows the single-neutron levels around the Fermi surface in the canonical basis for $^{40,42,44}\mathrm{Mg}$.
In $^{42}\mathrm{Mg}$, the gap between the $5/2^-$ level at $-2.45$ MeV and the $1/2^-$ level at $-0.44$ MeV is $2.01$ MeV.
In $^{44}\mathrm{Mg}$, the gap between the $5/2^-$ level at $-2.85$ MeV and the $1/2^-$ level at $-0.91$ MeV is $1.94$ MeV.
Following Refs.~\cite{ZhouSG2010PRC,LiLL2012PRC}, the single-neutron levels in $^{42}\mathrm{Mg}$ and $^{44}\mathrm{Mg}$ can be naturally divided into two parts: the well-bound $5/2^-$ level and the below ones corresponding to the ``core" part, and the remaining levels corresponding to the ``halo" part.
To understand the microscopic structures of the $K^\pi=1^-$ states below 3 MeV in $^{42}\mathrm{Mg}$ and $^{44}\mathrm{Mg}$, the QRPA amplitudes in the quasiparticle basis $X_{\mu\nu}^i$ and $Y_{\mu\nu}^i$ of these states are extracted from the QFAM amplitudes $X_{\mu\nu}(\omega)$ and $Y_{\mu\nu}(\omega)$ using the method from Ref.~\cite{Hinohara2013PRC}.
These amplitudes are then transformed into the canonical basis to get the amplitudes $X_{ab}^i$ and $Y_{ab}^i$ with $a,\,b$ denoting the canonical states.
After the summation of the indices $b$, the quantity $\sum_b(|X_{ab}^i|^2-|Y_{ab}^i|^2)$ measures the contribution of transitions from the single-particle level labeled by $a$.
Within this framework, the dominant transition for each state can be determined, and is schematically labeled in Fig.~\ref{fig3}.
It is found that the state at 2.50 MeV in $^{42}\mathrm{Mg}$ is predominantly (59.7\%) generated by such canonical states $ab$: the index $a$ corresponds to the $1/2^-$ single-neutron level at $-0.44$ MeV, and $b$ corresponds to various single-neutron levels with $\varepsilon>0\ \mathrm{MeV}$.
In other words, this specific state is dominated by transitions from the $1/2^-$ single-neutron level.
Similarly, the state at 2.56 MeV in $^{42}\mathrm{Mg}$ is dominated by transitions from the $3/2^-$ single-neutron level at $-0.22$ MeV (65.2\%);
the state at 2.30 MeV in $^{44}\mathrm{Mg}$ is dominated by transitions from the $1/2^-$ single-neutron level at $-0.91$ MeV (59.5\%);
the state at 2.39 MeV in $^{44}\mathrm{Mg}$ is dominated by transitions from the $3/2^-$ single-neutron level at $-0.86$ MeV (68.5\%).
In short, the $K^\pi=1^-$ states below 3 MeV in $^{42}\mathrm{Mg}$ and $^{44}\mathrm{Mg}$ are dominated by transitions from the ``halo" part of the single-neutron levels.

\begin{figure}[!htbp]
  \centering
  \includegraphics[width=0.8\textwidth]{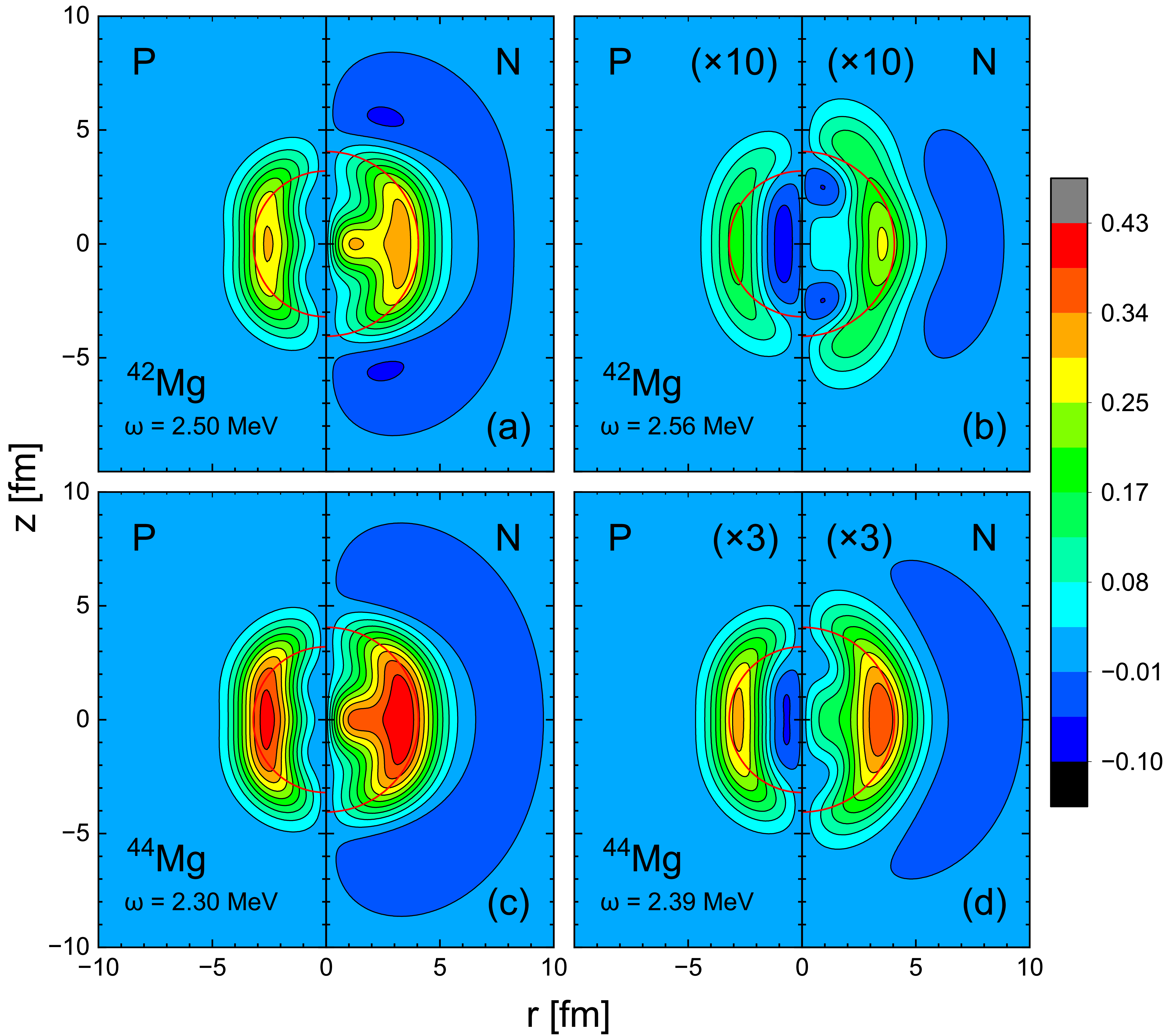}
  \caption{Transition densities of the $K^\pi=1^-$ states at 2.50 MeV (a) and 2.56 MeV (b) in $^{42}\mathrm{Mg}$, 2.30 MeV (c) and 2.39 MeV (d) in $^{44}\mathrm{Mg}$, calculated by QFAM with a Lorentzian smearing width of 0.002 MeV. In each panel the left part and right part correspond to the proton and neutron, respectively. The solid red lines indicate the root-mean-square radii of neutron and proton densities predicted by the DRHBc theory. The transition densities in (b) and (d) have been multiplied by 10 and 3, respectively.}
  \label{fig4}
\end{figure}

In Fig.~\ref{fig4}, the transition densities, defined as the imaginary part of the induced density, i.e., $\rho_{\mathrm{tr}}(\bm r)=\mathrm{Im}\delta j^0(\bm r)$, for $K^\pi=1^-$ states at 2.50 MeV and 2.59 MeV in $^{42}\mathrm{Mg}$, and at 2.30 MeV and 2.39 MeV in $^{44}\mathrm{Mg}$ are shown to understand the spatial structure of these states.
Around the surface and in the interior of the nucleus, neutrons and protons oscillate predominantly in phase.
Outside the nucleus, only neutrons oscillate, and the oscillation of these neutrons and the core is out of phase.
Combined with the analysis in Fig.~\ref{fig3}, the transition densities show that these $K^\pi=1^-$ states correspond to low-frequency oscillations of the neutron halo against the core in $^{42}\mathrm{Mg}$ and $^{44}\mathrm{Mg}$.
These results provide a microscopic picture for the soft dipole resonance in deformed halo nuclei $^{42}\mathrm{Mg}$ and $^{44}\mathrm{Mg}$.

\section{Summary}\label{summary}

We have developed the quasiparticle finite amplitude method (QFAM) based on the deformed relativistic Hartree-Bogoliubov theory in continuum (DRHBc) for multipole excitations, by explicitly linearizing the Dirac Hamiltonian and pairing potential.
As an example, the isovector electric dipole responses in neutron-rich even-even magnesium isotopes $^{28-44}\mathrm{Mg}$ are studied, focusing on the predicted deformed halo nuclei $^{42}\mathrm{Mg}$ and $^{44}\mathrm{Mg}$.

Systematic calculations of the isovector electric dipole strength functions in $^{28-44}\mathrm{Mg}$ reveal that the deformation splits the $K^\pi=0^-$ and $K^\pi=1^-$ giant dipole resonances.
For energies below 6 MeV,  from $^{28}\mathrm{Mg}$ to $^{40}\mathrm{Mg}$, there is no significant discrepancy between $K^\pi=0^-$ and $K^\pi=1^-$ strength functions.
While in $^{42}\mathrm{Mg}$ and $^{44}\mathrm{Mg}$, the strength for $K^\pi=1^-$ is much more enhanced than that of $K^\pi=0^-$.
Further analysis reveals that compared to $^{40}\mathrm{Mg}$,  the strength functions in $^{42}\mathrm{Mg}$ and $^{44}\mathrm{Mg}$ show additional $K^\pi=1^-$ peaks at 2.50 MeV and 2.30 MeV, respectively, involving contributions from two $K^\pi=1^-$ states.

The analysis of the QRPA amplitudes in the canonical basis shows that these $K^\pi=1^-$ states are dominated by transitions from the ``halo" part of the single-neutron levels, specifically the $1/2^-$ and $3/2^-$ levels near the Fermi surface.
The transition densities further reveal that around the surface and in the interior of the nucleus, neutrons and protons oscillate predominantly in phase.
While outside the nucleus, only neutrons oscillate, and the oscillation of these neutrons and the core is out of phase.
This spatial structure of the transition density, together with the dominant transitions from halo single-neutron states, provides a microscopic picture for the soft dipole resonance in $^{42}\mathrm{Mg}$ and $^{44}\mathrm{Mg}$.

\begin{acknowledgments}
  X.F.J thanks N. Hinohara, Y. F. Niu, C. Pan, H. Sagawa, K. Yoshida, K. Y. Zhang, and P. W. Zhao for stimulating discussions.
  Helpful suggestions from the members of the DRHBc Mass Table Collaboration are highly appreciated.
  This work was partly supported by the National Natural Science Foundation of China (Grants No. 12435006, No. 12475117, No. 12141501, and No. 12405132), the National Key R\&D Program of China (Contracts No. 2024YFE0109803 and No. 2024YFA1612600), the State Key Laboratory of Nuclear Physics and Technology, Peking University (Grant No. NPT2025KFY02), the National Key Laboratory of Neutron Science and Technology (Grants No. NST202401016), and the High-performance Computing Platform of Peking University.
\end{acknowledgments}

\end{document}